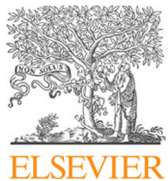
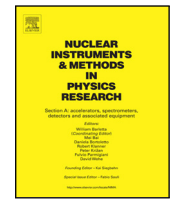

# Calorimeter development for the SuperNEMO double beta decay experiment

A.S. Barabash [a], A. Basharina-Freshville [b,c,*], S. Blot [c], M. Bongrand [d], Ch. Bourgeois [d], D. Breton [d], V. Brudanin [e,f], H. Burešovà [g], J. Busto [h], A.J. Caffrey [i], S. Calvez [d], M. Cascella [b], C. Cerna [j], J.P. Cesar [k], E. Chauveau [j,*], A. Chopra [b], G. Claverie [j], S. De Capua [c], F. Delalee [j], D. Duchesneau [l], V. Egorov [e], G. Eurin [d,b], J.J. Evans [c], L. Fajt [m], D. Filosofov [e], R. Flack [b], X. Garrido [d], H. Gómez [d], B. Guillon [o], P. Guzowski [c], R. Hodák [m], K. Holý [n], A. Huber [j], C. Hugon [j], A. Jeremie [l], S. Jullian [d], M. Kauer [b], A. Klimenko [e], O. Kochetov [e], S.I. Konovalov [a], V. Kovalenko [e], K. Lang [k], Y. Lemière [o], T. Le Noblet [l], Z. Liptak [k], X.R. Liu [b], P. Loaiza [d], G. Lutter [j], J. Maalmi [d], M. Macko [n,m,j], F. Mamedov [m], C. Marquet [j,*], F. Mauger [o], I. Moreau [j], B. Morgan [p], J. Mott [b], I. Nemchenok [e], M. Nomachi [q], F. Nova [k], H. Ohsumi [r], R.B. Pahlka [k], J.R. Pater [c], F. Perrot [j], F. Piquemal [j,s], P. Povinec [n], P. Přidal [m], Y.A. Ramachers [p], A. Rebii [j], A. Remoto [l], B. Richards [b], J.S. Ricol [j], C.L. Riddle [i], E. Rukhadze [m], R. Saakyan [b], R. Salazar [k], X. Sarazin [d], J. Sedgbeer [t], Yu. Shitov [e,t], F. Šimkovic [n], L. Simard [d,u], A. Smetana [m], K. Smolek [m], A. Smolnikov [e], S. Snow [p], S. Söldner-Rembold [c], B. Soulé [j], M. Špavorová [m], I. Štekl [m], J. Thomas [b], V. Timkin [e], S. Torre [b], Vl.I. Tretyak [v], V.I. Tretyak [e], V.I. Umatov [a], C. Vilela [b], V. Vorobel [w], D. Waters [b], A. Žukauskas [w]

[a] *NRC "Kurchatov Institute", ITEP, 117218 Moscow, Russia*
[b] *University College London, London WC1E 6BT, United Kingdom*
[c] *University of Manchester, Manchester M13 9PL, United Kingdom*
[d] *LAL, Université Paris-Sud, CNRS/IN2P3, Université Paris-Saclay, F-91405 Orsay, France*
[e] *JINR, 141980 Dubna, Russia*
[f] *National Research Nuclear University MEPhI, 115409 Moscow, Russia*
[g] *Nuvia a.s., Třebíč, Czech Republic*
[h] *CPPM, Université d'Aix-Marseille, CNRS/IN2P3, F-13288 Marseille, France*
[i] *Idaho National Laboratory, Idaho Falls, ID 83415, USA*
[j] *Université de Bordeaux, CNRS/IN2P3, CENBG, F-33175 Gradignan, France*
[k] *University of Texas at Austin, Austin, TX 78712, USA*
[l] *LAPP, Université Savoie Mont-Blanc, CNRS/IN2P3, F-74941 Annecy-le-Vieux, France*
[m] *Institute of Experimental and Applied Physics, Czech Technical University in Prague, CZ-12800 Prague, Czech Republic*
[n] *FMFI, Comenius University, SK-842 48 Bratislava, Slovakia*
[o] *LPC Caen, ENSICAEN, Université de Caen, CNRS/IN2P3, F-14050 Caen, France*
[p] *University of Warwick, Coventry CV4 7AL, United Kingdom*
[q] *Osaka University, 1-1 Machikaney arna Toyonaka, Osaka 560-0043, Japan*
[r] *Saga University, Saga 840-8502, Japan*
[s] *Laboratoire Souterrain de Modane, F-73500 Modane, France*
[t] *Imperial College London, London SW7 2AZ, United Kingdom*
[u] *Institut Universitaire de France, F-75005 Paris, France*
[v] *Institute for Nuclear Research, MSP 03680, Kyiv, Ukraine*
[w] *Charles University in Prague, Faculty of Mathematics and Physics, CZ-12116 Prague, Czech Republic*

## ARTICLE INFO

## ABSTRACT

SuperNEMO is a double-$\beta$ decay experiment, which will employ the successful tracker–calorimeter technique used in the recently completed NEMO-3 experiment. SuperNEMO will implement 100 kg of double-$\beta$ decay

* Corresponding authors.
*E-mail addresses:* anastasia.freshville@ucl.ac.uk (A. Basharina-Freshville), chauveau@cenbg.in2p3.fr (E. Chauveau), marquet@cenbg.in2p3.fr (C. Marquet).

Scintillators
Photomultipliers
Double beta decay
SuperNEMO

isotope, reaching a sensitivity to the neutrinoless double-$\beta$ decay ($0\nu\beta\beta$) half-life of the order of $10^{26}$ yr, corresponding to a Majorana neutrino mass of 50–100 meV. One of the main goals and challenges of the SuperNEMO detector development programme has been to reach a calorimeter energy resolution, $\Delta E/E$, around $3\%/\sqrt{E(\text{MeV})}\ \sigma$, or $7\%/\sqrt{E(\text{MeV})}$ FWHM (full width at half maximum), using a calorimeter composed of large volume plastic scintillator blocks coupled to photomultiplier tubes. We describe the R&D programme and the final design of the SuperNEMO calorimeter that has met this challenging goal.

## 1. Introduction

The SuperNEMO detector design is based on the technology of the recently completed NEMO-3 experiment [1–3], using a tracker–calorimeter detection technique to study neutrinoless double-$\beta$ decay ($0\nu\beta\beta$). It is a detector with multi-observables that allow full topological reconstruction of events resulting in powerful background rejection. The SuperNEMO detector will hold 100 kg of double-$\beta$ decay isotope ($^{82}$Se is the 'baseline' design isotope, with other isotopes being considered depending on enrichment possibilities) to reach a sensitivity of the order of $10^{26}$ years to the half-life of $0\nu\beta\beta$ decay, corresponding to a 50–100 meV effective Majorana neutrino mass [4]. A dominant factor in achieving the target sensitivity is the product $\Delta E \times N_{bkg}$, where $\Delta E$ is the energy window of the $0\nu\beta\beta$ decay at the $Q$ value of the decay, $Q_{\beta\beta}$, approximated by the energy resolution of the detector in keV, and $N_{bkg}$ is the expected background index in kg$^{-1}$keV$^{-1}$yr$^{-1}$. Consequently the two main areas of focus for the SuperNEMO calorimeter R&D programme were resolution and radiopurity. SuperNEMO's event topology reconstruction capabilities will help suppress backgrounds from natural radioactivity. However, the double-$\beta$ decay with neutrino emission ($2\nu\beta\beta$), allowed in the Standard Model, will have the same topological signature as most $0\nu\beta\beta$ mechanisms. The tail of the continuum spectrum from the summed electron energy distribution of the $2\nu\beta\beta$ decay may extend into the energy window near the $Q_{\beta\beta}$ value making it an irreducible background to the $0\nu\beta\beta$ process. Improving the energy resolution is the only way to reduce this background and is the focus of the work described here.

The SuperNEMO detector has a modular design (Fig. 1), where the detector and the double-$\beta$ decay source are distinct. The detector consists of 20 modules, each being 4 m in height, 6 m in length and 2 m in width. One module contains 5 kg of $^{82}$Se in the form of a thin ($\approx 40$ mg/cm$^2$) vertically suspended source foil, surrounded by 2000 drift cells operating in Geiger mode (for particle tracking) and enclosed by calorimeter walls consisting of 520 optical modules (for energy and time of flight measurements). Each optical module is a square-faced scintillator block coupled to an 8-in. photomultiplier tube (PMT). The SuperNEMO detection principle is shown in Fig. 2.

A current-carrying coil wrapped around the module produces a magnetic field of 25 G to distinguish electrons from positrons. Passive shielding will surround the detector to reduce the environmental neutron and $\gamma$-ray background. The construction of the first SuperNEMO module, known as the Demonstrator, is nearing its end with first data expected towards the end of 2017.

The main requirements of the SuperNEMO calorimeter are to provide a good energy and time resolution for low energy ($\mathcal{O}(1$ MeV)) electron detection as well as efficient $\gamma$-ray tagging ($>50\%$ at 1 MeV) for background suppression. The SuperNEMO calorimeter must be optimised to detect incoming electrons simultaneously originating from the same vertex in the double-$\beta$ decay source foils (Fig. 2). The calorimeter also needs to use robust and long-lasting technology that is easy to manufacture and assemble whilst considering the channel count and the cost. Taking into account these requirements and radiopurity constraints, a scintillator-based detector is the optimal choice for SuperNEMO.

Given the constraints of using a scintillator detector and the goal of reaching a sensitivity of the order of $10^{26}$ years for an exposure of 500 kg yr, the requirement for the SuperNEMO calorimeter energy resolution ($\Delta E/E$) for electrons is set to be around $7\%/\sqrt{E(\text{MeV})}$ FWHM (full width at half maximum), or 4% FWHM at 3 MeV, the $Q_{\beta\beta}$ value of $^{82}$Se. This can be seen in Fig. 3, which shows simulations for the SuperNEMO half-life sensitivity as a function of the calorimeter energy resolution for a fixed exposure of 500 kg yr. The $\Delta E/E$ required for SuperNEMO represents a factor of two improvement over the energy resolution of the NEMO-3 calorimeter, which was $(14-17)\%/\sqrt{E}$ (MeV) for electrons, despite the NEMO-3 scintillator blocks being 50% smaller in volume than those of SuperNEMO.

This paper will discuss the SuperNEMO calorimeter requirements and the parameters that influence the energy resolution of a scintillator-PMT optical module (Section 2), the calorimeter test bench used to test the various parameters (Section 3), and the final SuperNEMO calorimeter design and the R&D programme carried out to reach it (Section 4).

## 2. SuperNEMO calorimeter requirements

The calorimeter R&D programme for SuperNEMO has covered four main areas of study: geometry, energy resolution, radiopurity and calibration.

### 2.1. Geometry

The calorimeter is divided into two walls on either side of the double-$\beta$ source foil, as shown in Fig. 1. It must be segmented in order to measure the individual energy of each particle, whilst minimising the dead zones between each optical module and having a reasonable number of channels. The thickness of the scintillator blocks must be at least 3 cm to fully absorb the electrons produced in $0\nu\beta\beta$, and greater than 10 cm for efficient $\gamma$-ray background identification. The calorimeter geometry must also take into account the necessity to use a magnetic shield around the PMT to shield it from the 25 G magnetic field.

### 2.2. Energy resolution

The energy resolution of scintillator detectors, $\Delta E/E$, is dominated by stochastic photoelectron fluctuations. If the number of photoelectrons is large, as is the case for the SuperNEMO calorimeter, a Gaussian approximation can be used for the energy resolution:

$$\frac{\Delta E}{E} = \frac{2.35\sigma}{E} = \frac{2.35}{\sqrt{N_{pe}}}, \qquad (2.1)$$

where $\sigma$ is the width of the Gaussian function. The number of photoelectrons ($N_{pe}$) can be written as a function of the deposited energy $E$ and of terms related to the scintillator properties, the light collection by the PMT and the intrinsic properties of the PMT:

$$N_{pe} = E\,[\text{MeV}] \cdot N_{ph}^{0} \cdot \epsilon_{col}^{light} \cdot (QE \cdot \epsilon_{col}^{\text{PMT}}). \qquad (2.2)$$

Here, $N_{ph}^{0}$ is the number of scintillation photons per 1 MeV of deposited energy and is determined by the scintillator light output. The term $\epsilon_{col}^{light}$ is the collection efficiency of the scintillation light by the PMT;

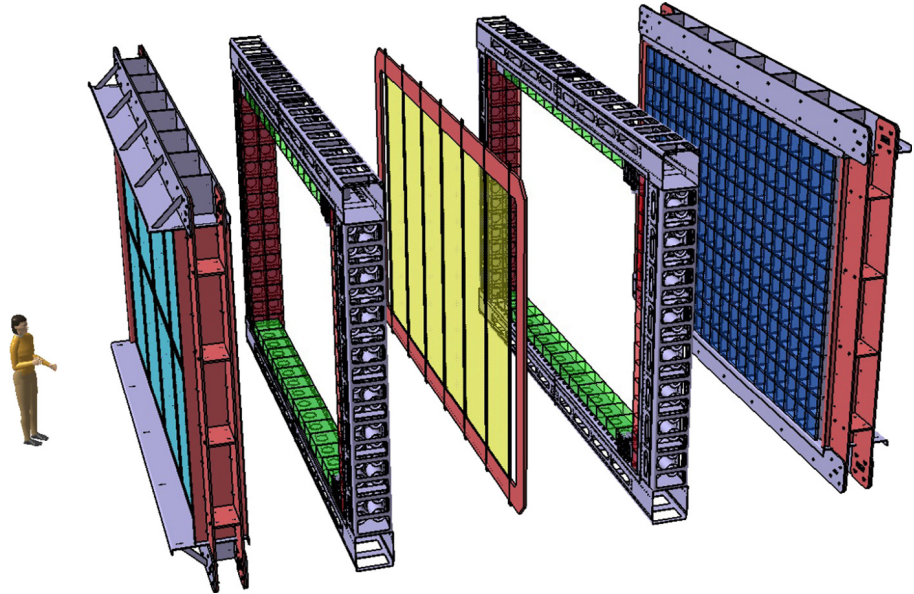

**Fig. 1.** An exploded view of a SuperNEMO module, showing (from left to right) one calorimeter wall, one tracker volume, the source foil, another tracker volume and another calorimeter wall.

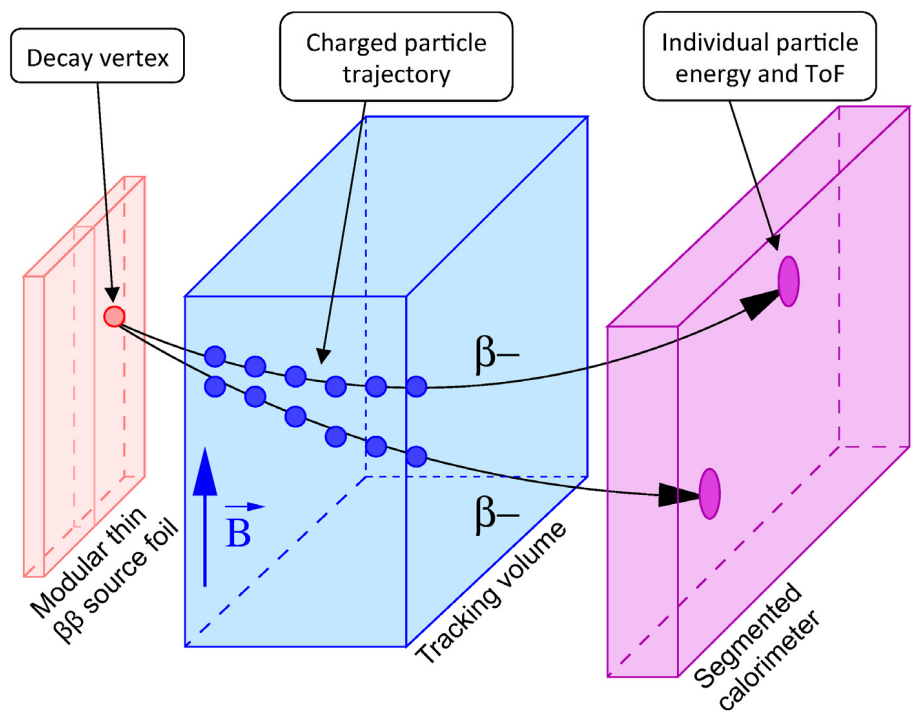

**Fig. 2.** The SuperNEMO detection principle.

it depends on the material, geometry and surface treatment of the scintillator. It is also a function of the reflector material efficiency, in which the optical module is wrapped, and of the optical coupling quality between the scintillator and the PMT. The quantum efficiency of the photocathode, $QE$, is one of the critical intrinsic characteristics of the PMT along with the collection efficiency $\epsilon_{col}^{PMT}$ of photoelectrons from the photocathode to the dynode system. In order to achieve the target sensitivity, the energy resolution of the SuperNEMO calorimeter is required to be around 7% FWHM at 1 MeV for electrons. This energy resolution corresponds to a target of at least 1100 photoelectrons per MeV as can be seen from Eq. (2.1), which requires optimisation of the parameters corresponding to the terms described in Eq. (2.2).

The choices of scintillator material and PMT type for the SuperNEMO calorimeter are discussed in Sections 4.1 and 4.2, respectively.

Good uniformity is required within a scintillator block, with the energy resolution at any point on the entrance face of the block not varying more than ±10% relative to the energy resolution obtained at the central position of the block. This requirement ensures that the non-uniformity of an individual scintillator block does not exceed the spread in the energy response from different blocks introduced by the mass production of optical modules. The block's response can also be corrected using the spatial resolution of the SuperNEMO tracking detector, which is $\sigma_L$ = 13 mm and $\sigma_T$ = 0.7 mm [5], where $\sigma_L$ is the resolution along the wire direction and $\sigma_T$ is the resolution perpendicular to it.

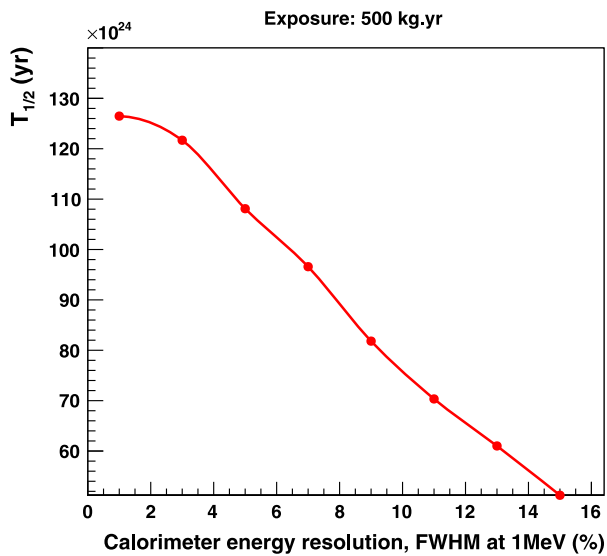

**Fig. 3.** SuperNEMO half-life sensitivity as a function of energy resolution with a fixed exposure of 500 kg yr for a $^{82}$Se source foil thickness of 40 mg/cm$^2$ and the SuperNEMO target foil contamination of 2 μBq/kg for $^{208}$Tl, 10 μBq/kg for $^{214}$Bi and 0.15 mBq/m$^3$ for $^{222}$Rn in the tracking volume [4].

Based on NEMO-3 experience [3] and the larger size of the scintillator blocks in SuperNEMO compared to that of NEMO-3, the requirement for the time resolution is set to be $\sigma_t = 400$ ps at 1 MeV between two calorimeter optical modules in coincidence for rejection of background external to the foil. This is required to discriminate between two-electron events (the signature for $0\nu\beta\beta$) originating in the source foils from those that originate outside of the detector and then cross its active volume to imitate $0\nu\beta\beta$ events. Improving the time resolution of the calorimeter is not the purpose of the R&D programme, however it has benefited from the high light output achieved to meet the energy resolution goals. The time resolution of the optical modules has been monitored at every stage of the R&D programme.

### 2.3. Radiopurity

Due to the low background requirements of the experiment, ultra-radiopure materials must be used throughout the detector, paying particular attention to $^{40}$K, $^{214}$Bi and $^{208}$Tl, which can affect the background level and the counting rate. The activity levels of radioisotopes of the plastic scintillators selected for SuperNEMO (2.2 ± 1 mBq/kg for $^{40}$K, <0.3 mBq/kg for $^{214}$Bi and <0.1 mBq/kg for $^{208}$Tl) are negligible compared to the PMTs, and in particular the PMT glass, which are the main source of contamination. The PMT radiopurity requirements depend on the double-$\beta$ decay isotope being studied (the lower the $Q_{\beta\beta}$ value the more stringent the requirements) and are at the level of 150 mBq/kg for $^{40}$K, 65 mBq/kg for $^{214}$Bi and 4 mBq/kg for $^{208}$Tl for double-$\beta$ decay isotopes with a $Q_{\beta\beta}$ value at 3 MeV.

### 2.4. Calibration

During the five years of planned SuperNEMO data taking, the gain and stability of a large number of PMTs must be monitored at the 1% level. The detector energy response must be linear, and any non-linear effects must be controlled at the 1% level up to electron energies of 3–4 MeV, which is the region of interest for $0\nu\beta\beta$. The absolute calibration system will use $^{207}$Bi sources to provide K-shell conversion electrons with energies of 482 keV, 976 keV and 1682 keV, inserted into the detector at monthly intervals. A light injection calibration system based on ultraviolet light-emitting diodes (UV LED) will be used for daily monitoring and correction of the PMT gain drift, as well as to check the linearity of the calorimeter response. Using low activity $\alpha$ sources embedded into the scintillator to monitor the gain is an additional possibility which has been studied and will be tested on some of the optical modules that will be used in the LED calibration system for the SuperNEMO Demonstrator. The embedded alpha sources consist of an $^{241}$Am deposit sandwiched between two 100 μm scintillator plates that are optically glued onto the main scintillator block. A $^{60}$Co source, providing two coincident $\gamma$-rays of 1.17 and 1.33 MeV, will occasionally be used for absolute time calibration. The SuperNEMO calorimeter calibration will follow a similar procedure to that employed in NEMO-3 [1]. It is not discussed further in this work.

### 3. Calorimeter test bench

The energy resolution measurement is carried out on optical modules of different configurations. The measurement is carried out by exciting the scintillator with a flux of mono-energetic electrons.

Two electron sources are used. The main one, used for detailed characterisation, is a $^{90}$Sr based electron beam passing through a magnetic field to select electrons of a narrow energy (FWHM = 1.0±0.2% at 1 MeV). Background is suppressed by a coincidence trigger module consisting of a 130 μm plastic scintillator placed in the electron beam upstream of the optical module under test [6]. Passing the electron beam through this trigger module introduces an additional 0.5% into the $\Delta E/E$ measurement due to the fluctuation of energy losses of the electrons in the thin scintillator. This contribution is quadratically subtracted from the final $\Delta E/E$ result obtained with a SuperNEMO optical module. The beam can be moved across the face of the scintillator block to measure the uniformity of the response. The energy of the beam can be varied from 0.4 to 1.8 MeV by changing the magnetic field settings. This feature is used to measure the linearity of the optical module response and the dependence of the energy resolution on the beam energy. In addition, a $^{207}$Bi conversion electron source is used as a cross-check measurement [7].

The calorimeter time resolution is measured by a 420 nm LED signal delivered to two optical modules in coincidence. The LED is driven by a pulser to create a signal shape similar to that produced by ionising radiation events in the scintillator. The reference time resolution is measured at the 1 MeV equivalent electron energy deposition. This LED system is also used to measure the linearity of the optical modules [8].

The test bench measurements have been compared with GEANT-4 based Monte Carlo (MC) optical simulations. These simulations model light emission, propagation and collection inside an optical module, taking into account wavelength dependence of all of the optical processes for the specific characteristics of the optical module material (light yield, refractive index, attenuation length, reflectivity, quantum efficiency and photoelectron collection). The optical simulations and the experimental data are within a 2% agreement [9].

### 4. Optical module design

Many different configurations of optical modules have been tested to determine the final design of the SuperNEMO calorimeter, which will consist of polystyrene (PS) plastic 256 mm × 256 mm square scintillator blocks with a minimal thickness of 141 mm and a hemispherical 'cutout' directly coupled to an 8-in. PMT (Fig. 4) covered by a magnetic shield.

In order to increase the light collection, the blocks are wrapped in 600 μm Teflon® on the sides followed by 12 μm aluminised Mylar® on all the faces, apart from the hemispherical cutout for the PMT. The following sections describe the R&D process that has led to this design.

#### 4.1. Scintillators

Considering dimensional constraints imposed by the design of SuperNEMO and the requirement of having a manageable number of channels in the detector, four main scintillator geometries were considered. These are described in Table 1. Scintillator characteristics were studied and compared by carrying out measurements using identical PMTs for each set of tests.

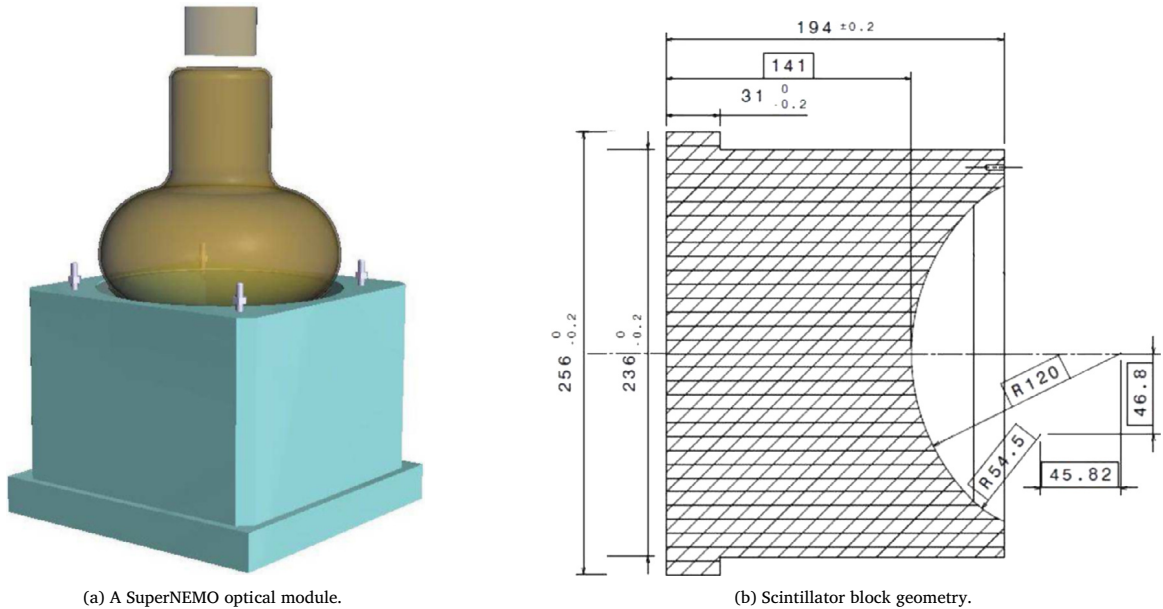

(a) A SuperNEMO optical module.  (b) Scintillator block geometry.

**Fig. 4.** The SuperNEMO calorimeter optical module: a 256 mm × 256 mm scintillator block coupled to an 8-in. PMT.

**Table 1**
SuperNEMO calorimeter candidate scintillator geometries. C stands for cuboid, H for hexagonal and T for tapered shapes.

| Geometry name | Geometry type | Entrance face dimensions (mm) | Depth (mm) |
|---|---|---|---|
| C256 | | 256 × 256 | 190 |
| C308 | | 308 × 308 | 190 |
| T308 | | 308 × 308 | 190 |
| H276 | | 276 ⌀ | 120 |

### 4.1.1. Material

The calorimeter requires a scintillator that has a high light yield, low electron backscattering, which is proportional to $Z^2$, high radiopurity, good timing and a relatively low cost. A fast inorganic option of YSO ($Y_2SiO_5$ doped with Ce) and Phoswich consisting of two scintillators ($CaF_2$ doped with Eu and poly(vinyltoluene)) optically coupled to each other were considered but have not met the requirements of radiopurity, backscattering and cost. The most obvious type of scintillator that is able to satisfy all of these requirements simultaneously is organic scintillator. Initially, liquid (toluene-based) scintillating agent and liquid-plastic hybrid scintillators were considered for the calorimeter [10]. However, due to difficulties with the mechanical design and degradation of $\Delta E/E$ as a result of the energy loss of electrons in the entrance window of the liquid scintillator container these options were not considered for the final design.

The two main choices of low-Z scintillator are polystyrene-based as used for the NEMO-3 detector, and poly(vinyltoluene)-based (PVT). These scintillators are composed of scintillating and wavelength shifting agents, the composition and concentration of which are not disclosed by most manufacturers. The scintillator manufacturer options for SuperNEMO were JINR Dubna, ISM Kharkiv and NUVIA CZ for PS, Saint-Gobain Crystals and Detectors, and Eljen Technology for PVT, with the characteristics of the main candidates listed in Table 2. Whilst the PS and PVT scintillators had similar nominal performances to each other for each respective type, the two main manufacturers considered were NUVIA CZ for PS and Eljen Technology for PVT due to the range of products available and R&D opportunities, as well as their mass production capabilities.

PVT has a higher light yield than PS and therefore is expected to provide a better energy resolution. However, PVT presents some mechanical challenges, such as 'crazing' of the scintillator surfaces, which occurs when it comes into contact with various common substances. PVT is also more brittle than PS and hence more difficult to machine. The cost of PVT is also higher than PS, therefore both the PVT and PS options have been considered for the R&D. A comparison of measurements made with PVT and PS scintillators for a cubic block of $308 \times 308 \times 190$ mm$^3$, named C308, is shown in Table 3, confirming that PVT gives a better $\Delta E/E$ with an improvement factor, $f_{\text{FWHM}}$, of 1.14 for EJ-204 scintillator relative to the JINR Dubna PS scintillator block. This number is consistent with the expected light yield ratio (Table 2).

An R&D programme was undertaken with NUVIA CZ to improve the performance of their PS. The improvements consisted of cleaner conditions for the scintillator production procedure and a refined concentration of the scintillating and wavelength shifting agents. This improved scintillator, known as enhanced PS, has a composition of 1.5% p-Terphenyl (p-TP) and 0.05% POPOP (1.4-bis(5-phenyloxazol-2-yl) benzene) and shows an improvement factor, $f_{\text{FWHM}}$, of ($1.04 \pm 0.03$) for $\Delta E/E$ relative to the standard NUVIA CZ PS production [11], measured with a $256 \times 256 \times 190$ mm$^3$ cubic scintillator block, named C256 (Table 4).

### 4.1.2. Geometry and optical coupling

Several geometries have been closely studied, including the geometries of a basic square (C256, C308), a tapered square (T308) and a tapered hexagon (H276). The dimensions of the entrance face are optimised to fit the calorimeter wall and the practical considerations on the number of channels. Tapered geometries have been considered in order to reduce the amount of material and thus, to test the possible effects on the length of photon trajectories. The hexagonal shape was designed to get closer to a cylindrical shape in order to limit edge effects on light propagation. MC simulations and measurements were carried out, leading to a choice of a hexagonal H276 or square C256 face scintillator (Fig. 5) [12]. The tapered T308 geometry did not show any improvement due to the tapered face sending light back towards the entrance face of the block. As the H276 and C256 geometries have the same entrance face area and give similar $\Delta E/E$ results, the square block has been chosen for the final design due to ease of manufacturing and construction.

Table 2
SuperNEMO calorimeter candidate scintillators and their commercial characteristics, where $\lambda$ is the wavelength of maximum emission (nm).

| Material | Type | Light yield (ph/1 MeV $e^-$) | $\lambda$ (nm) | Refr. index at 589.3 nm/$\lambda$ | Decay Time (ns) | Atten. Length (cm) |
|---|---|---|---|---|---|---|
| PS | NUVIA CZ | ≈ 8,500 | 425 | 1.57/1.60 | 2.5 | Not available |
| PS | JINR NEMO-3 | ≈ 8,000 | 430 | 1.57/1.60 | 2.5 | 200 |
| PS | ISM Kharkiv | ≈ 8,000 | 430 | 1.57/1.60 | 2.5 | 200 |
| PVT | Saint-Gobain BC-404 | ≈ 10,400 | 408 | 1.58/1.61 | 1.8 | 140 |
| PVT | Saint-Gobain BC-408 | ≈ 10,000 | 425 | 1.58/1.61 | 2.1 | 210 |
| PVT | Eljen Technology EJ-204 | ≈ 10,400 | 408 | 1.58/1.61 | 1.8 | 160 |
| PVT | Eljen Technology EJ-200 | ≈ 10,000 | 425 | 1.58/1.61 | 2.1 | 380 |

Table 3
Summary of tests of PVT and PS scintillators, using a C308 geometry block and an 8-in. Photonis XP1866 PMT. The fraction $f_{FWHM}$ is the improvement factor relative to the JINR Dubna PS scintillator block.

| Material | $\Delta E/E$(%) | $f_{FWHM}$ |
|---|---|---|
| JINR NEMO-3 PS | 8.9 ± 0.2 | 1 |
| Eljen Technology EJ-200 PVT | 8.3 ± 0.2 | 1.07 ± 0.03 |
| Eljen Technology EJ-204 PVT | 7.8 ± 0.2 | 1.14 ± 0.03 |

Table 4
Summary of tests of NUVIA CZ and enhanced NUVIA CZ PS scintillators, using a C256 geometry block and an 8-in. Hamamatsu R5912-MOD PMT. The fraction $f_{FWHM}$ is the improvement factor relative to the NUVIA CZ PS scintillator block.

| Material | $\Delta E/E$(%) | $f_{FWHM}$ |
|---|---|---|
| NUVIA CZ PS | 7.9 ± 0.2 | 1 |
| Enhanced NUVIA CZ PS | 7.6 ± 0.2 | 1.04 ± 0.03 |

Table 5
Summary of tests of optical coupling materials, using an Eljen Technology EJ-200 PVT scintillator of a C308 geometry and an 8-in. Hamamatsu R5912-MOD PMT, where $f_{FWHM}$ is an improvement factor relative to the isopropanol alcohol.

| Optical material | Refractive index | $\Delta E/E$ (%) | $f_{FWHM}$ |
|---|---|---|---|
| Isopropanol alcohol | 1.37 | 9.4 ± 0.2 | 1 |
| Cargille gel | 1.46 | 8.6 ± 0.2 | 1.09 ± 0.04 |
| Cargille gel | 1.52 | 8.4 ± 0.2 | 1.12 ± 0.04 |
| RTV 615 | 1.41 | 9.4 ± 0.2 | 1.00 ± 0.03 |

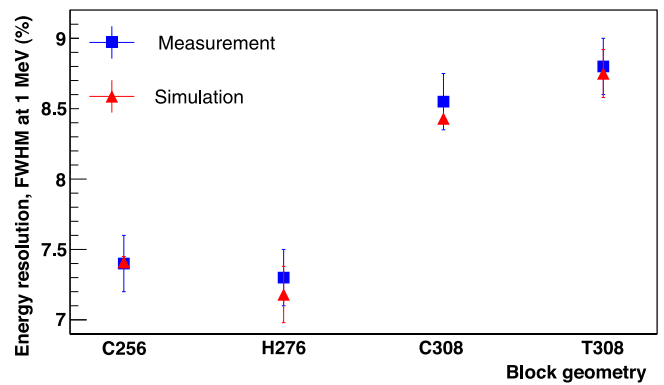

Fig. 5. Energy resolution, $\Delta E/E$, measured (blue squares) and simulated (red triangles) for 1 MeV electrons for different geometries of PVT scintillator blocks coupled to an 8-in. Hamamatsu R5912-MOD PMT. (For interpretation of the references to colour in this figure legend, the reader is referred to the web version of this article.)

Due to the presence of a ≈25 G magnetic field in SuperNEMO, the PMTs need to be protected with a pure iron magnetic shield. In order to achieve the optimal packing fraction of individual C256 scintillator blocks while allowing for space for the magnetic shield around each optical module, the scintillator block has a 31 mm deep 'step' extending 10 mm around the main body of the block, as shown in Fig. 4(b). The thickness of the step is chosen to fully absorb the electrons incident on it, where 30 mm is required for electrons up to 3 MeV from the baseline $^{82}$Se isotope. The length of the shield required to protect the optical module from the magnetic field (see Section 4.2.2 and Fig. 8(a) for further details) constrains the distance between the scintillator front face and the PMT photocathode to be 141 mm (Fig. 4(b)).

Comparisons of optical modules with and without lightguides coupling the scintillator to the PMT have been made. Introducing a lightguide into the setup increases the number of optical contacts in the module, which leads to poorer light collection and hence a worse $\Delta E/E$. Measurements with and without lightguides were carried out on a NEMO-3 square face block (200 mm × 200 mm × 100 mm) coupled to a 5-in. Hamamatsu R6594 PMT. Removing the lightguide from the setup gives a significant improvement factor $f_{FWHM} = (1.20 \pm 0.02)$ for $\Delta E/E$ with respect to coupling via a lightguide.

The hemispherical 8-in. PMT is therefore directly coupled to the scintillator, which has a hemispherical cutout. The depth of the cutout has been optimised considering the photocathode size, uniformity, the ease of gluing and the possibility of dismantling the PMT from the scintillator block. Taking into account these parameters, measurements have been carried out to find the cutout depth corresponding to the optimal polar angle of the coverage of the PMT photocathode by the scintillator block. This angle was found to be 80° (0° corresponds to the PMT axis) and gives an improvement factor $f_{FWHM} = (1.05 \pm 0.03)$ for $\Delta E/E$ compared to the initial coverage of 60°. Further increasing the depth does not result in any $\Delta E/E$ improvement due to the size of the photocathode and a smaller quantum efficiency at the photocathode edges.

Good optical coupling between the scintillator and PMT is essential for uniform and complete light collection, mostly achieved by coupling the PMT directly to the scintillator, as described above. However, the coupling material used between the PMT and scintillator can also provide a sizeable contribution. It should have a refractive index in-between those of the scintillator (1.58) and the PMT glass (1.47). Its transparency, radiopurity, viscosity and bonding properties also have to be considered. The viscosity needed is such that the glue does not leak out from the scintillator-PMT interface before it is set during the assembly of optical modules. Tests have been carried out with optical gels of different optical indices (from Cargille Cie), showing an improvement factor $f_{FWHM} = (1.12 \pm 0.04)$ in $\Delta E/E$ for a gel with a refractive index of 1.52 (Table 5) with respect to isopropanol.[1] Glues with this refractive index exist, however none of them have been found to satisfy the requirements placed on bonding and radiopurity—these glues are transparent to radon emanation and would therefore be a concern for background contribution.

---

[1] Isopropanol was used as a coupling material for all of the short-term tests to ease the testing procedure.

Table 6
The main SuperNEMO calorimeter candidate PMTs and their characteristics for the best selected tubes.

| Type | Cathode diameter (in.) | Quoted Max QE | QE at 400 (nm) | Dynode stages | Gain at nominal HV |
|---|---|---|---|---|---|
| Hamamatsu R5912-MOD | 8 | 42% | 41% | 8 | $5 \times 10^5$ at 1500 V |
| Photonis XP1886 | 8 | 35% | 35% | 8 | $5 \times 10^5$ at 1500 V |
| ET Enterprises 9354 kB | 8 | 28% | 26% | 12 | $7 \times 10^6$ at 1300 V |

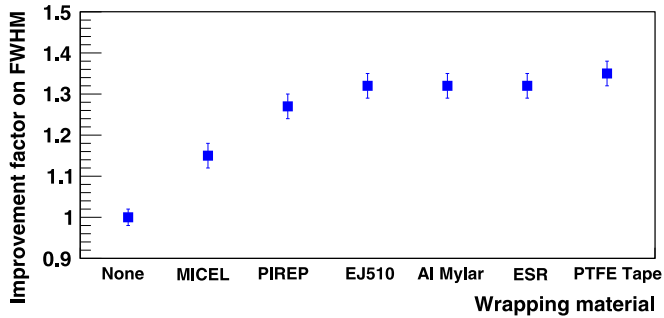

Fig. 6. Summary of tests of specular and diffusive reflective materials for the wrapping of PS C308 block sides (with the entrance face wrapped in aluminised Mylar®) coupled to an 8-in. R5912-MOD Hamamatsu PMT.

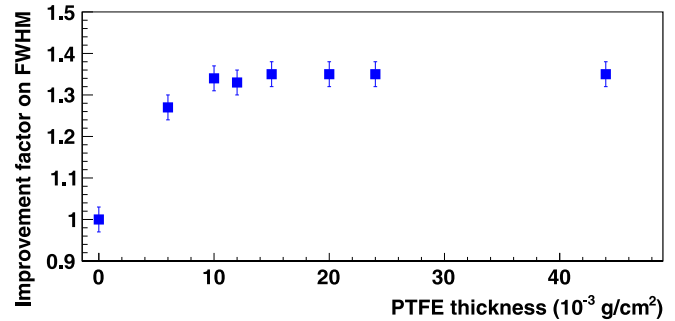

Fig. 7. Summary of tests of different thicknesses of Teflon® (PTFE) tape from GEB for the wrapping of PS C308 block sides (with the entrance face wrapped in aluminised Mylar®) coupled to an 8-in. R5912-MOD Hamamatsu PMT.

An optical epoxy silicone rubber compound RTV 615, with a refractive index of 1.41, is currently to be used to couple the PMTs to the blocks in the construction of SuperNEMO. The compound has a lower refractive index than needed, but has high radiopurity and the required viscosity. It is therefore used as a compromise between the various optical coupling requirements.

*4.1.3. Wrapping and surface finishing*

In order to increase light collection, the optical modules must be wrapped in a reflective material, which must have high reflectivity to redirect any escaped light back to the PMT, and be radiopure. In addition, the reflective material on the scintillator entrance face must have a low Z to reduce electron backscattering. It must also be dimensionally thin to minimise energy loss of electrons as they pass through the material, but thick enough to shield the optical module from the UV photons produced in the SuperNEMO tracker. Its thickness must also be known to a high precision to have a good understanding of the energy losses of electrons crossing it. These criteria led to the choice of 12 μm of aluminised Mylar® for the entrance face of the block. Various reflectors have been tested to determine the best wrapping for the remaining faces. The $\Delta E/E$ can depend on whether such reflectors are specular or diffusive. Diffusively reflecting Teflon® (PTFE) tape (from GEB), sheets (50 μm thick from MICEL and 500 μm from PIREP), and paint (Eljen Technology EJ-510) have been compared against Enhanced Specular Reflector (ESR) (Fig. 6). Optical simulations and measurements have shown that the best $\Delta E/E$ is achieved using diffusive reflectors on the sides of the block with thicknesses greater than 0.010 g/cm² (Fig. 7) [13]. However, to ensure the reliability of the wrapping three layers of PTFE tape with a total thickness of 600 μm, or 0.024 g/cm², are used. The back face of the block is covered by the 12 μm of aluminised Mylar®, which encloses the entrance face as well as the side faces making for an easier wrapping procedure.

Studies have also been carried out to see what kind of surface finishing would get the best performance from the optical module, leading to all surfaces of the scintillator being of a raw finish, except for the hemispherical cutout surface, which will be polished. The polishing of the cutout surface leads to an improvement factor $f_{\text{FWHM}} = (1.08 \pm 0.03)$ on the $\Delta E/E$.

*4.2. Photomultiplier tubes*

The SuperNEMO calorimeter requires a PMT with a high QE, good photoelectron collection efficiency, gain that provides a linear response, high radiopurity, good time resolution and low dark current. Different size PMTs were considered for SuperNEMO, starting with a 5-in. PMT since anything smaller would not be viable due to the large number of channels and cost, and going up to an 11-in. PMT. During testing it was found that using a 5-in. PMT did not achieve the required $\Delta E/E$ as its size did not match the geometry and size of the chosen scintillator blocks. The 10-in. and 11-in. PMTs did not achieve the required $\Delta E/E$ due to an insufficient collection efficiency from the photocathode to the first dynode, which becomes smaller for larger diameter PMT bulbs. This led to a choice of an 8-in. PMT, which, relative to NEMO-3, increases the photo-detection surface to improve the $\Delta E/E$ and decreases the total number of calorimeter channels used.

Most of the R&D tasks described could only be carried out in partnership with PMT manufacturer companies, prompting the SuperNEMO Collaboration to start close work with Photonis and Hamamatsu in 2005. These companies started the development of 8-in. PMTs (XP1886 from Photonis and R5912-MOD from Hamamatsu, Table 6). ET Enterprises PMTs were also considered and tested, however they did not have the QE required for SuperNEMO (Table 6).

*4.2.1. Photocathode quantum efficiency*

The development of new photocathode processes in recent years has been achieved with bi-alkali alloys, such as SbKCs and SbKNa [14], leading to a peak QE of about 40%, initially for 3-in. PMTs. These photocathodes have a spectral sensitivity optimal in the UV to blue region, thus are a good match for the peak emission wavelength ($\lambda$) of the SuperNEMO scintillators (Table 2). Photonis and Hamamatsu have worked on extrapolating these processes to 8-in. PMTs and have produced several tubes with QE at or above ≈35% at about 400 nm (Table 6). For comparison, the average QE of the 3-in. and 5-in. PMTs used in NEMO-3 detector was 25%.

The QE value depends on the wavelength and on the location of the light incident on the photocathode surface. Measurements have shown that the QE is constant over almost the entire surface of the Photonis PMTs. For Hamamatsu PMTs, the QE sharply decreases from 80% to 0% of its maximal value for the photon impact point at angles greater than 70° (0° corresponds to the PMT axis). After further development by Hamamatsu, the photocathode uniformity was improved to being within 95% of its maximal value across the entire surface of the PMT, which improved the $\Delta E/E$ by a factor $(1.06 \pm 0.04)$ (Table 7).

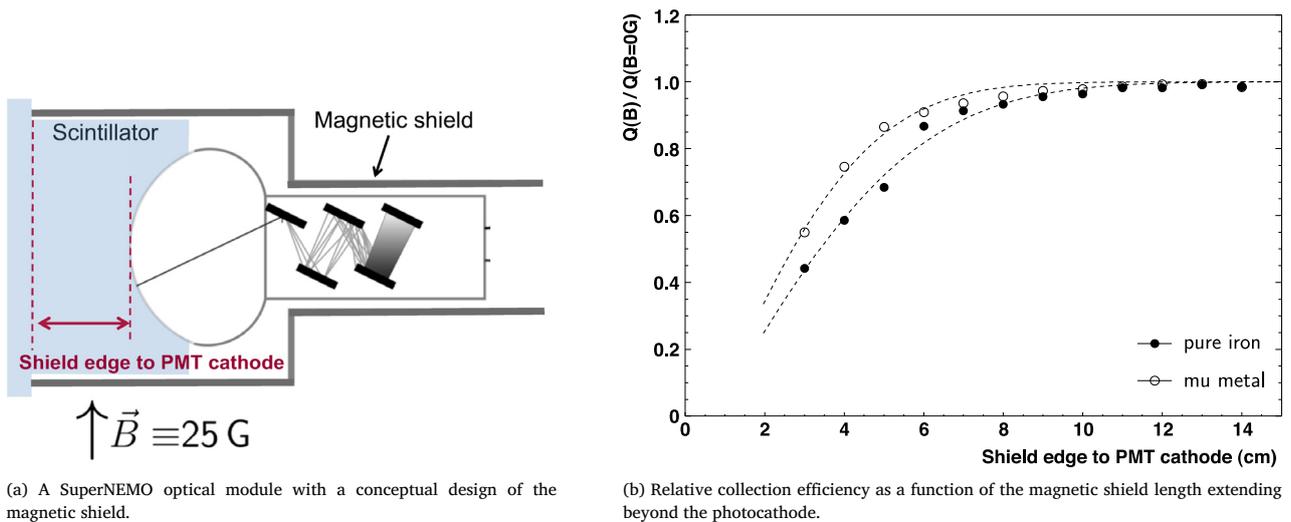

(a) A SuperNEMO optical module with a conceptual design of the magnetic shield.

(b) Relative collection efficiency as a function of the magnetic shield length extending beyond the photocathode.

**Fig. 8.** A SuperNEMO optical module (a) and relative collection efficiency (b) as a function of distance to protect the PMT photocathode inside a 25 G magnetic field.

**Table 7**
Summary of tests of Hamamatsu R5912-MOD PMTs before and after photocathode uniformity improvements, using a C256 Eljen Technology EJ-204 scintillator block, where BI indicates the PMT before improvement and AI indicates the PMT after improvement, run at the nominal high voltage. $f_{\text{FWHM}}$ is an improvement factor relative to the R5912-MOD BI tube.

| PMT | $\Delta E/E$ (%) | $f_{\text{FWHM}}$ | Charge (pC) | Rise time (ns) |
| --- | --- | --- | --- | --- |
| R5912-MOD BI | 8.0 ± 0.2 | 1 | 98 ± 2 | 4.1 ± 0.3 |
| R5912-MOD AI | 7.5 ± 0.2 | 1.06 ± 0.04 | 108 ± 2 | 4.4 ± 0.3 |

*4.2.2. Photoelectron collection efficiency*

Although the QE of the Hamamatsu tubes is higher (Table 6), the Photonis tubes were initially found to give a better $\Delta E/E$. R&D carried out by Hamamatsu on the vacuum properties of the tubes improved the collection efficiency and therefore the $\Delta E/E$. The High Voltage (HV) distribution of the first two stages of the voltage divider was optimised to improve the collection efficiency of the photoelectrons. This was achieved by increasing the HV between the cathode and the first dynode by a factor of 2 and between the first and second dynodes by a factor of 1.5, resulting in a relative $\Delta E/E$ improvement factor $f_{\text{FWHM}} = (1.05 \pm 0.03)$ [13].

The impact of SuperNEMO's 25 G magnetic field on the photoelectron collection efficiency has also been studied. Measurements have been carried out using a dedicated test bench to guide the design of the magnetic shield for an optical module. The optical module under test is placed in a solenoid providing a magnetic field of 25 G and its response to injected light from a UV LED is measured. The PMT signal charge is recorded as a function of the magnetic shield coverage of the photocathode, as illustrated in Fig. 8(a). The result shown in Fig. 8(b) demonstrates that the magnetic shield should extend at least 10 cm beyond the PMT photocathode to maintain the charge collection efficiency. Based on these results, 3 mm thick annealed pure iron has been chosen for the magnetic shield that encloses the optical module and covers a 110 mm length of the scintillator in front of the PMT entrance face. As previously shown, these requirements for the magnetic field have strongly impacted on the scintillator geometry design.

*4.2.3. Gain and linearity*

The 8-in. and larger diameter PMTs have so far been mostly used for Cerenkov light detection, offering a large surface and high gain ($10^8$–$10^{10}$) to detect the few photoelectron output. In SuperNEMO the level of light is expected to be about 1100 photoelectrons at 1 MeV, corresponding to an instantaneous peak current of about 3 mA. Therefore, non-linearity effects on the energy response have been considered in the design of 8-in. PMTs. The 10 or 11 dynode stages commonly used in large PMTs have been replaced by 8 dynode stages with a focused linear geometry of the dynodes in order to reduce the gain to around $10^6$. The electric field has been increased progressively between the last dynodes to prevent any space charge effects. The distribution of HV between dynodes is obtained with a voltage divider circuit and the optimal ratio between resistors of the divider has been found to be 20–6–4–1–1.25–1.5–1.75–2–2 from the photocathode to the anode (initially, the ratio was 10–4–4–1–1–1–1–1–1). Larger resistor values between the first two dynodes also help improve the photoelectron collection efficiency, as discussed in Section 4.2.2.

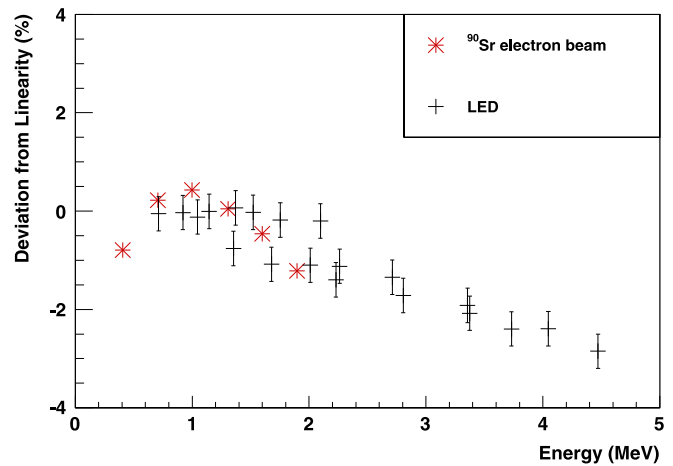

**Fig. 9.** Observed deviation from a linear fit of the charge response of a SuperNEMO optical module (a C256 NUVIA CZ PS block and an 8′ Hamamatsu R5912-MOD PMT) as a function of electron energy from the $^{90}$Sr based electron beam and electron energy equivalent using a LED.

Measurements carried out with a LED tuned to reproduce the light level in a SuperNEMO optical module corresponding to an energy range of 0–4.5 MeV (Fig. 9) show that the deviation from linearity does not exceed 1.5% at 3 MeV.

*4.2.4. Time resolution*

The scintillator and the PMT each account for half of the time resolution of the optical module. In the scintillator, the propagation of the photons dominates over the emission time of the scintillating agent and wavelength shifter. Reducing the number of dynode stages of the

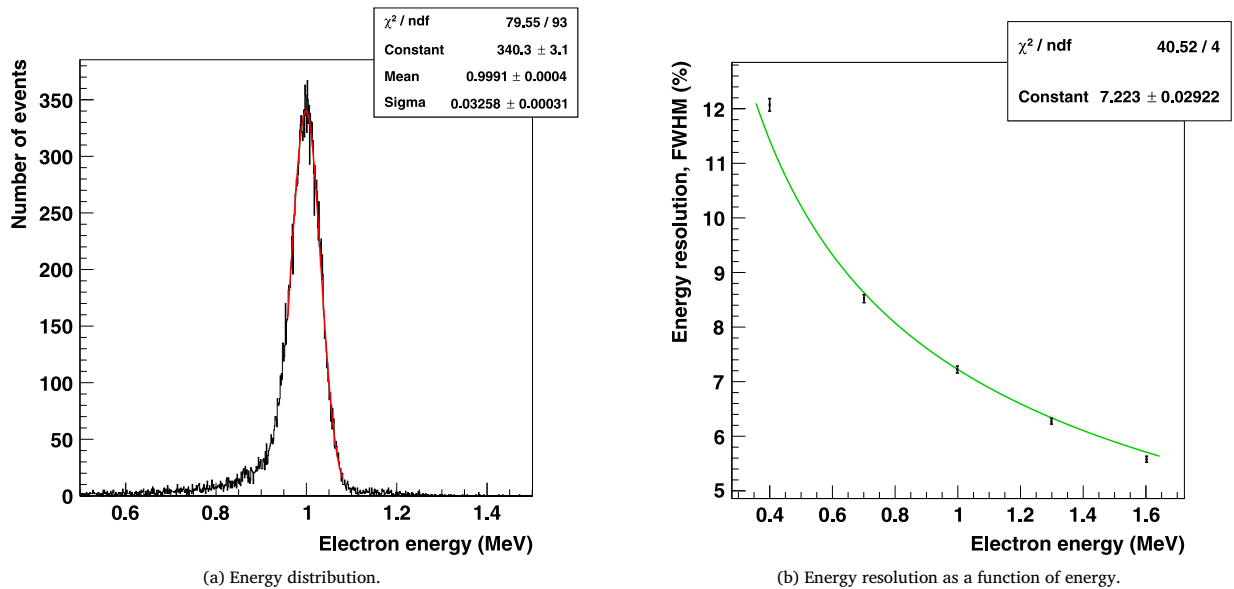

(a) Energy distribution.

(b) Energy resolution as a function of energy.

**Fig. 10.** Measurements for a C256 Eljen Technology PVT block coupled to an 8-in. Hamamatsu R5912-MOD PMT: (a) The energy distribution measured with 1 MeV electrons fit with a Gaussian function. This distribution includes an additional 0.5% $\Delta E/E$ caused by the coincidence trigger module scintillator used in the $^{90}$Sr based electron beam (Section 3). (b) Energy resolution as a function of the energy of the $^{90}$Sr electron beam fit with a Constant/$\sqrt{E}$ function.

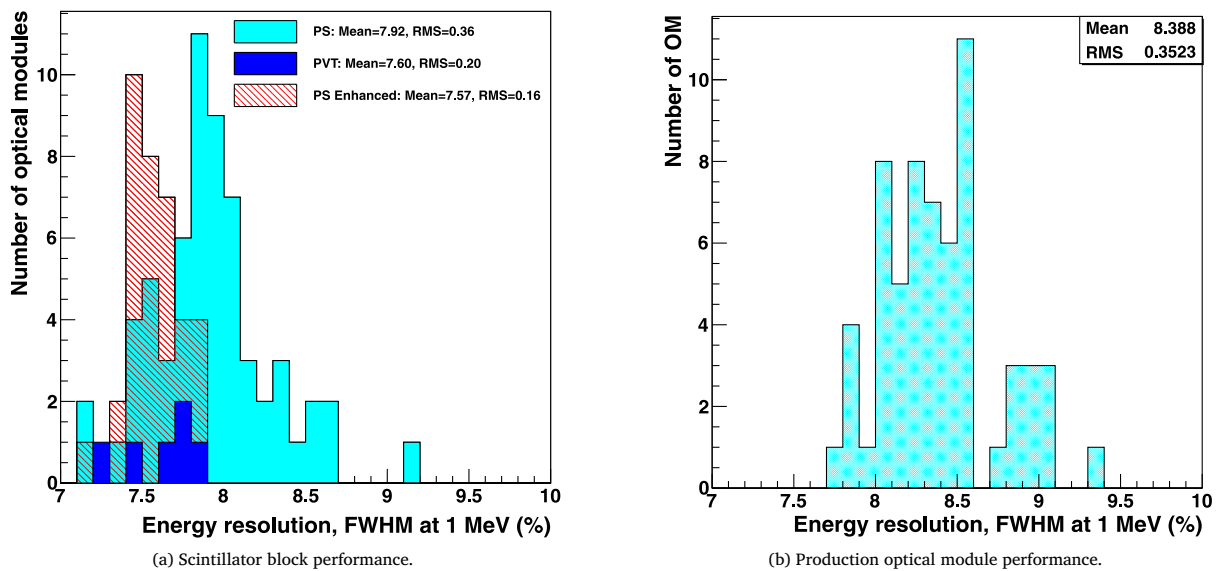

(a) Scintillator block performance.

(b) Production optical module performance.

**Fig. 11.** (a) Scintillator performance: Energy resolution of 6 PVT, 62 PS and 36 Enhanced PS C256 blocks obtained with the same 8-in. Hamamatsu R5912-MOD PMT. (b) Production performance: Energy resolution of 62 optical modules (a C256 PS block coupled to an individual 8-in. Hamamatsu R5912-MOD PMT) produced for the SuperNEMO Demonstrator.

PMT helped reduce the transit time spread (TTS) of the PMT, improving the timing of the calorimeter. The hemispherical shape of the PMT also ensures better timing uniformity by equalising the photoelectron trajectories inside the PMT.

The main focus of the calorimeter development has been on optimising the light collection for $\Delta E/E$, therefore the time resolution has only been validated for the configurations that have achieved the required $\Delta E/E$. It has been measured with LED light delivered to two optical modules in coincidence. This time resolution measured for a 1 MeV signal is equal to $(400 \pm 90)$ ps [8].

### 4.2.5. Afterpulses and dark noise

PMT afterpulses could lead to an increased counting rate of the calorimeter and false event identification, which can affect the background models and introduce systematics in signal selection. If included in the charge integration of pulses, they could also degrade the $\Delta E/E$. The PMTs undergo a selection criterion where the probability of afterpulses occurring within 2 μs of the main pulse should be <1%. All PMTs have passed this selection.

Due to the high light yield of SuperNEMO optical modules (1100 photoelectrons at 1 MeV) a dark noise rate of <5 Hz at a threshold of 5 photoelectrons will provide a negligible contribution to the detector counting rate. This level of dark noise rate is easily achievable by most manufacturers of 8-in. PMTs.

### 4.2.6. Radiopurity

To ensure radiopurity, all the components of the optical module, particularly the PMT components and glass, are selected using high purity Germanium (HPGe) detectors. The glass of the PMT is the main source of contamination in the calorimeter. The current level of radiopurity reached for the glass of Hamamatsu 8-in. PMTs is about 850 mBq/kg for $^{40}$K, 380 mBq/kg for $^{214}$Bi and 150 mBq/kg for $^{208}$Tl, to be compared with the analogous NEMO-3 5-in. PMT values of 1400 mBq/kg for $^{40}$K, 650 mBq/kg for $^{214}$Bi and 40 mBq/kg for $^{208}$Tl [1]. These are at the level of the requirements reported in Section 2.3 and are sufficient to reach the projected sensitivity with the SuperNEMO Demonstrator module. R&D is currently ongoing to make further radiopurity improvements in order to accommodate the multi-isotope strategy of the full SuperNEMO detector, in particular for isotopes with lower $Q_{\beta\beta}$ values.

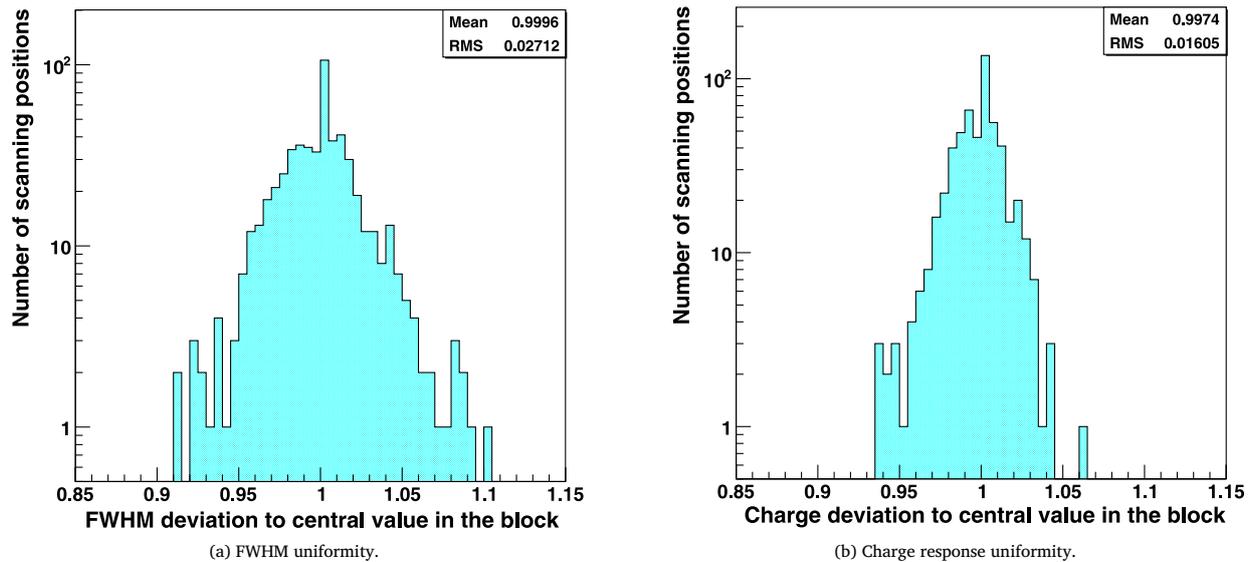

(a) FWHM uniformity.

(b) Charge response uniformity.

**Fig. 12.** The uniformity of the scintillator response across the entrance face of C256 blocks measured at 9 positions for 62 blocks for FWHM (a) and the response in charge (b). The data are taken with the $^{90}$Sr electron beam at 9 different points across the scintillator face and then normalised to the central value.

*4.3. Achieved performances*

The best energy resolution has been achieved for an optical module with a PVT C256 scintillator block directly coupled using RTV 615 to an 8-in. Hamamatsu R5912-MOD PMT with a polished hemispherical cutout in the scintillator. Similar performances have been achieved with an enhanced PS C256 scintillator block, as well as with a Photonis XP1886 8-in. PMT. The production of the latter was, however, discontinued in 2009. The block is wrapped in 600 μm Teflon® on the sides and 12 μm aluminised Mylar® on all sides and faces of the scintillator, apart from the hemispherical cutout for the PMT, to increase light collection. This configuration gives a $\Delta E/E$ result of $(7.2 \pm 0.2)$% for 1 MeV electrons. The energy measurement of 1 MeV and the dependence of energy resolution on the $^{90}$Sr based electron beam energy, which follows a $1/\sqrt{E}$ distribution, are shown in Fig. 10.

The SuperNEMO modules will use the 8-in. R5912-MOD Hamamatsu PMTs. For cost effectiveness NUVIA CZ PS scintillator has been chosen. Both the standard and enhanced NUVIA CZ productions have been selected for the Demonstrator in order to compare their performances relative to each other on a mass production scale. The scintillator performance is shown in Fig. 11(a), which gives the distribution of $\Delta E/E$ at 1 MeV with 6 Eljen Technology PVT, 62 standard NUVIA CZ PS and 36 enhanced NUVIA CZ PS scintillator blocks, all obtained with the C256 geometry and the same 8-in. R5912-MOD Hamamatsu PMT. Fig. 11(b) shows the energy resolution of the optical modules produced for the SuperNEMO Demonstrator. In order to show the spread introduced by the PMTs the same 62 standard PS blocks as those shown in Fig. 11(a) are chosen, but this time each coupled to an individual PMT in accordance with the optical module production procedure.

The spatial uniformity of the optical module in terms of the impact point of the electrons on the entrance face of the scintillator has been measured for 9 positions on each block with an electron beam size of a 3 mm Gaussian width. The face of the scintillator is divided into 9 equal squares and the energy response and the $\Delta E/E$ are measured at the centre of each of these squares. The spatial distribution of the $\Delta E/E$ is within 10% of the value measured at the centre of the block for 98% of the optical modules (Fig. 12(a)) and within 5% for 96% of the optical modules for the charge response (Fig. 12(b)).

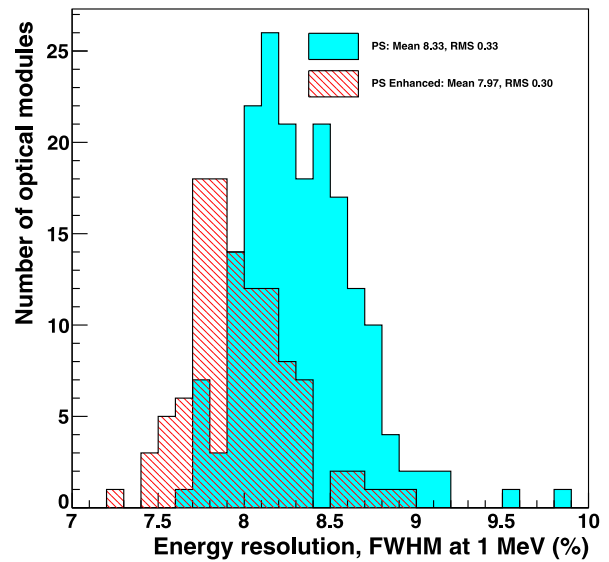

**Fig. 13.** Production performance of SuperNEMO Demonstrator optical modules, showing the energy resolution of 184 PS and 113 Enhanced composition PS C256 blocks coupled to an individual 8-in. Hamamatsu R5912-MOD PMT.

**5. Summary and conclusion**

The SuperNEMO experiment aims to reach a sensitivity to the half-life of the $0\nu\beta\beta$ decay of $^{82}$Se of the order of $10^{26}$ years with 100 kg of isotope in 5 years. One of the main focuses of the project R&D has been on the calorimeter design, in particular improving the energy resolution, FWHM, for 1 MeV electrons from (14–17)% achieved by NEMO-3 to around 7% for SuperNEMO using larger volume plastic scintillators without compromising other requirements such as the linearity in energy, time resolution, radiopurity, reliability and cost effectiveness. The optimal configuration consists of a 256 mm×256 mm×194 mm (C256) PS scintillator block, directly coupled via RTV 615 to an 8-in. Hamamatsu R5912-MOD PMT via a hemispherical cutout in the scintillator. The collaboration has achieved the $\Delta E/E$ stipulated by the R&D proposal with the best optical modules showing an energy resolution of 7.2% at 1 MeV. The mean energy resolution of the production batch modules is 8.3% at 1 MeV for the standard NUVIA CZ production PS and 8.0% at 1 MeV for the enhanced NUVIA CZ production PS (Fig. 13). The observed difference between the $\Delta E/E$ of the production and the R&D stipulation is well within the expected range due to tolerances of PMT and scintillator manufacturing processes.

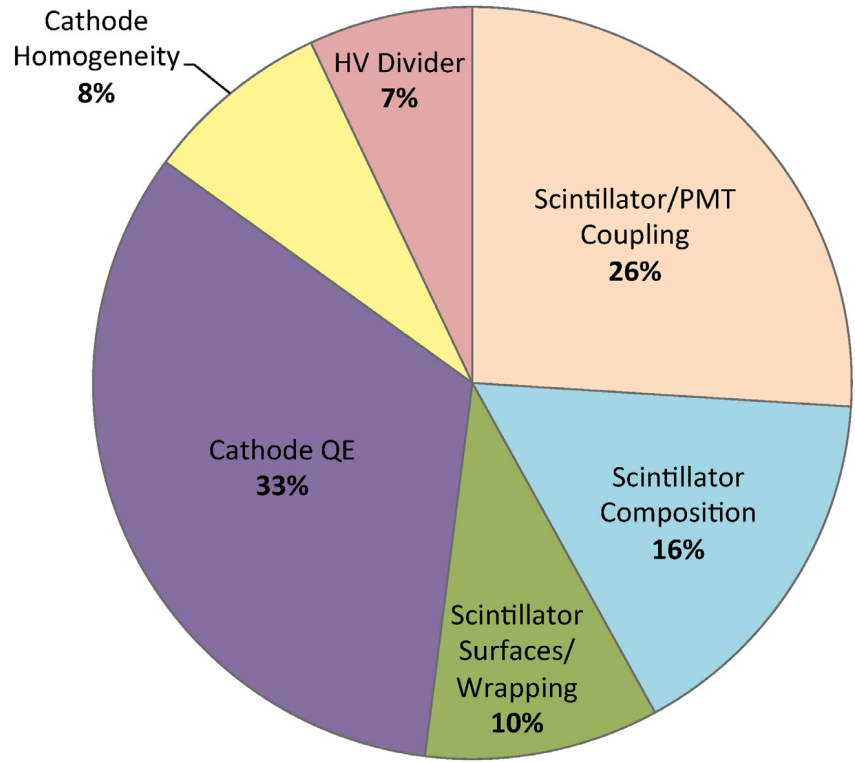

**Fig. 14.** Contribution of individual R&D program parts to the improvement in FWHM achieved with the SuperNEMO calorimeter compared to NEMO-3.

The NUVIA CZ enhanced scintillator production already employs cleaner conditions for the production, with further improvement in this area possible. Other improvements are also possible if a radiopure, transparent optical coupling agent for non-permanent bonding of the PMT to the scintillator block with a better matched refractive index is found. Further collaboration with PMT manufacturers aimed at providing a better match between the photocathode wavelength response and the emission spectra of the SuperNEMO scintillator blocks could also lead to an improvement in the $\Delta E/E$.

The contribution of each part of the R&D to the improvement in $\Delta E/E$ can be seen in Fig. 14, where the starting point is the $\Delta E/E$ of NEMO-3 (14–17)%/$\sqrt{E(\text{MeV})}$ and the endpoint is the $\Delta E/E$ of SuperNEMO 7.2%/$\sqrt{E(\text{MeV})}$. The largest improvement for $\Delta E/E$ comes from the increase in quantum efficiency of the bi-alkali photocathodes of the 8-in. PMTs. Another large contributing factor to the $\Delta E/E$ improvement is due to the direct coupling of the 8-in. PMT to the hemispherical cutout in the scintillator block without the use of a lightguide. Optimising the material, wrapping and surface finishing, geometry of the scintillator block, as well as operation of the PMT (optimisation of the voltage divider, gain and linearity) have also significantly contributed to reaching the target $\Delta E/E$.

The SuperNEMO Demonstrator module, which will host 7 kg of $^{82}$Se in the first instance, is currently being installed in the Laboratoire Souterrain de Modane (LSM) and will start data taking in 2017.

We acknowledge support by the grants agencies of the CNRS/IN2P3 in France, STFC in the UK, NSF in the USA, RFBR in Russia, and the Ministry of Education, Youth and Sports of the Czech Republic.